\documentclass[12pt]{article}
\usepackage{latexsym}
\def\bseq{\begin{subequation}}  % = 1a 1b
\def\eseq{\end{subequation}}
\def\bsea{\begin{subeqnarray}}  % = 1.1a 1.1b
\def\esea{\end{subeqnarray}}
\newcommand{\bbox}{\lower.2ex\hbox{$\Box$}}

\newcommand{\beq}{\begin{equation}}
\newcommand{\eeq}{\end{equation}}
\newcommand{\bea}{\begin{eqnarray}}
\newcommand{\eea}{\end{eqnarray}}
\newcommand{\ena}{\end{eqnarray}}

\renewcommand{\a}{\alpha}
\renewcommand{\b}{\beta}

\renewcommand{\d}{\delta}

\newcommand{\pa}{\partial}
\newcommand{\g}{\gamma}
\newcommand{\G}{\Gamma}

\newcommand{\D}{\Delta}

\newcommand{\m}{\mu}

\newcommand{\n}{\nu}

\newcommand{\p}{\pi}

\newcommand{\s}{\sigma}

\newcommand{\ad}{{\dot{\alpha}}}

\begin{document}
\begin{titlepage}
\begin{flushright}
IFUM-FT-671\\

%hep-th/0012009
\end{flushright}
\vspace{2cm}
%\begin{center}
\noindent{\Large \bf Noncommutative ${\cal N}=1,2$
 super $U(N)$ Yang-Mills:

\vspace{3mm}
\noindent UV/IR mixing and effective action results at one loop}
\vspace{1cm}
%\vfill
{\bf \hrule width 16.cm}%\vskip 15mm%27.mm
\vspace {1cm}
\noindent{\large \bf
Daniela Zanon}
\vskip 3mm%1cm
{ \small
\noindent Dipartimento di Fisica dell'Universit\`a di Milano
and

\noindent INFN, Sezione di Milano, Via Celoria 16,
20133 Milano, Italy}
%\end{center}
\vfill
\begin{center}
{\bf Abstract}
\end{center}
{\small Noncommutative ${\cal N}=1$ and  ${\cal N}=2$
supersymmetric Yang-Mills theories
with gauge group $U(N)$  are studied here using
the background field method and superspace background
covariant $D$-algebra in perturbation theory.  At
one loop divergences arise only in the
two-point functions. They are logarithmic UV/IR divergences
in the planar/nonplanar sectors that play a role dual to each other.

\noindent We  consider contributions
to the effective action with vector external lines.
We find that the three-point function vanishes, while the four-point function
receives contributions both from vector and from chiral matter loops. }
\vspace{2mm} \vfill \hrule width 6.cm
\begin{flushleft}
e-mail: daniela.zanon@mi.infn.it
\end{flushleft}
\end{titlepage}

Recently noncommutative geometry
received much attention both in string theory and in gauge field
theory \cite{all,SW,others,string,perturb}, and various aspects of the
intimate connection between the two approaches have been enlightened.

In this paper we restrict our attention to noncommutative field
theories
and continue the study of the effective action
for supersymmetric Yang-Mills theories in superspace
perturbation theory. In \cite{DZ,ASDZ} the theory under consideration has
been ${\cal N}=4$ supersymmetric Yang-Mills with
gauge group $U(N)$. It is the $4$-dimensional gauge field theory with the
largest number of symmetries and as such the simplest to be studied
even in its noncommutative version. The analysis is simplified if one
uses the superfield formulation \cite{FL} combined with the
background field method \cite{DZ}.
With these techniques perturbative calculations become manageable and
gauge invariance is kept under control. In \cite{DZ,ASDZ}
contributions to the one-loop effective action have been considered:
the two- and three-point  functions with
external background vector fields are zero as it happens in the commutative
case \cite{GS,GS2,superspace}.
In addition the four-point function has been computed and shown
to be in agreement with the field theory limit of corresponding string
amplitudes \cite{*trek}. The noncommutative theory is completely free
of ultraviolet and infrared divergences, as expected since the
corresponding commutative theory is conformally invariant. In the
background field method approach used in \cite{DZ,ASDZ} the
perturbative calculations of background vector $n$-point functions
are particularly simple because loops with quantum chiral superfields
need not be evaluated. It is a trivial matter to show that the three
ghost fields completely cancel corresponding contributions from the
three chiral matter superfields. After all ${\cal N}=4$ super
Yang-Mills has to be simple.

As soon as we relax the symmetry and look at the corresponding theories
with ${\cal N}=1$ and ${\cal N}=2$ supersymmetries the matter-ghost
cancellation no longer occurs and more ingenuity is required in order
to proceed and compute the one-loop higher-order functions. At
the level of the two-point function one finds UV and
IR divergences in the planar and in the non planar sectors
respectively. This is the UV/IR mixing that signals a $\b\neq 0$ in
the corresponding commutative theory.

Here we apply the same techniques introduced in \cite{DZ,ASDZ} to the
noncommutative ${\cal N}=1$ and ${\cal N}=2$ supersymmetric
Yang-Mills theories and consider corrections to the one-loop
effective action up to the four-point function.

\vspace{0.8cm}
The noncommutative ${\cal N}=1$ supersymmetric Yang-Mills theory is
described in ${\cal N}=1$ superspace by the following classical
action (with
notations and conventions as in \cite{superspace})
\beq
S= \frac{1}{4g^2}~{\rm Tr}\left. \left( \int d^4x~d^2\theta~ W^\a
W_\a
+ \int d^4x~d^2\bar{\theta}~ \bar{W}^\ad \bar{W}_\ad  \right)\right|_*
\label{N1SYMaction}
\eeq
where  $W^\a= i\bar{D}^2(e_*^{-V}*D^\a e_*^V)$ is the gauge
superfield strength. The fields are Lie-algebra valued, e.g.
$V=V_a T^a$, in the adjoint representation of $U(N)$.
In (\ref{N1SYMaction}) the symbol $|_*$ indicates that the
superfields are multiplied using  the $*$-product defined as
\beq
(\phi_1 * \phi_2)(x,\theta,\bar{\theta})\equiv e^{\frac{i}{2}
\Theta^{\m\n}\frac{\pa}{\pa x^\m}\frac{\pa}{\pa y^\n}}~
\phi_1(x,\theta,\bar{\theta})\phi_2(y,\theta,\bar{\theta})|_{y=x}
\label{starprod}
\eeq

If, in addition to the vector real superfield
$V$ we introduce a chiral matter superfield $\Phi$,
we can construct a theory in which
${\cal N}=2$ supersymmetry is realized. The corresponding classical
action describes  ${\cal N}=2$ noncommutative Yang-Mills
written in terms of ${\cal N}=1$ superfields
\beq
S= \frac{1}{g^2}~{\rm Tr}\left. \left( \int d^4x~d^4\theta~ e^{-V}
\bar{\Phi}e^{V} \Phi +\frac{1}{4}\int d^4x~d^2\theta~ W^\a W_\a
+\frac{1}{4} \int d^4x~d^2\bar{\theta}~
\bar{W}^\ad \bar{W}_\ad  \right)\right|_*
\label{N2SYMaction}
\eeq

The noncommutative theories in (\ref{N1SYMaction}) and  (\ref{N2SYMaction})
can be quantized using the background field method
\cite{DZ}, essentially adopting the same procedure as in the commutative
case \cite{GS,GZ}.

For the gauge multiplet one performs
a non linear splitting between
the quantum prepotential $V$ and the background superfield, via covariant
derivatives (in quantum chiral representation \cite{superspace})
\bea
&&\nabla_\a =e_*^{-V}* D_\a~ e_*^V
 ~\rightarrow~e_*^{-V}*\nabla^B_\a ~ e_*^V
\nonumber\\
&&~~~~\nonumber\\
&& \bar{\nabla}_\ad= \bar{D}_\ad
~\rightarrow~
 \bar{\nabla}^B_\ad
\label{backcovder}
\eea
In this way the external background enters in the quantum action implicitly
in the background covariant derivatives through the connections
\beq
\nabla^B_\a= D_\a-i{\bf{\G}}_\a \qquad \qquad \bar{\nabla}^B_\ad= \bar{D}_\ad-
i\bar{\bf{\G}}_\ad \qquad\qquad \nabla^B_a=\pa_a-i{\bf{\G}}_a
\label{backcovderconn}
\eeq
and explicitly in the background field strength
${\bf W}_\a=\frac{i}{2}[\bar{\nabla}^{B\ad},\{\bar{\nabla}^B_\ad,\nabla^B_\a\}]$.
Background covariant gauge-fixing introduces additional terms
\beq
-\frac{1}{4g^2} {\rm Tr} \int
d^4x~d^4\theta~V*(\nabla_B^2\bar{\nabla}_B^2+\bar{\nabla}_B^2\nabla_B^2)*V
+S_{FP}+S_{NK}
\label{gauge-fixing}
\eeq
with Faddeev-Popov action
\beq
S_{FP}= {\rm Tr}\int d^4x~d^4\theta~[\bar{c}'*c-c'*\bar{c}+\frac{1}{2}
(c'+\bar{c}')*[V,c+\bar{c}]_*+\dots]
\label{FP}
\eeq
and Nielsen-Kallosh ghost action
\beq
S_{NK}={\rm Tr}\int d^4x~d^4\theta~\bar{b}*{b}
\label{NK}
\eeq
In (\ref{FP}) we have used the notation $[A,B]_*=A*B-B*A$.
The three ghosts $c$, $c'$ and $b$ are background covariantly chiral
superfields, i.e. $\bar{\nabla}_B^\ad c=\bar{\nabla}_B^\ad c'
=\bar{\nabla}_B^\ad b=0$. While the Faddeev-Popov ghosts do have
interactions with the quantum $V$ fields, the Nielsen-Kallosh ghost
interacts only with the background. Therefore it contributes only at one
loop. In the following we drop the suffix $B$ from the covariant
derivatives since covariance is everywhere with respect to the background
and no ambiguity can arise.

After gauge-fixing the quadratic quantum $V$-action becomes
\beq
S\rightarrow -\frac{1}{4g^2} {\rm Tr} \int
d^4x~d^4\theta~V*(\Box-i{\bf W}^\a*\nabla_\a-i\bar{\bf W}^\ad*
\bar{\nabla}_\ad)*V
\label{actionquadratic}
\eeq
where $\Box\equiv1/2 \nabla^a\nabla_a$ is the background covariant
Laplacian.

For the matter superfield $\Phi$ which appear in (\ref{N2SYMaction})
the quantum-background splitting is a
standard linear one
\beq
\Phi~\rightarrow~\Phi+\Phi_B
\label{phiback}
\eeq
where $\Phi_B$ is the background external field. The superfields
$\Phi$ and $\Phi_B$ are background covariantly chiral
$\bar{\nabla}_\ad \Phi=0$.

Our final goal is to compute one-loop contributions to the effective actions
of ${\cal N}=1$ and ${\cal N}=2$ Yang-Mills with external vector
background fields. Therefore the corresponding actions quadratic in the
quantum fields are all we need for these one-loop
calculations. From (\ref{actionquadratic}) we see that the
interactions with the background
are at most linear in the $D$-spinor derivatives. Since at least two
$D$'s and two $\bar{D}$'s are needed in the loop in order to
complete the $D$-algebra, from (\ref{actionquadratic})
one starts obtaining a nonvanishing result
at the level of the four-point function.
We conclude that for the ${\cal N}=1$ theory the contributions to the vector
two- and three-point functions can
be generated only by ghosts circulating in the loop. For the ${\cal N}=2$
theory we have to take into account also the contributions from the
chiral matter loop.

 As mentioned above the
three ghosts are background covariant chiral superfields with actions
given in (\ref{FP}) and (\ref{NK}) respectively. Since we only
compute up to one loop, in (\ref{FP}) we can drop terms that contain
interactions with the quantum $V$-field. The same considerations apply
also to the chiral superfield $\Phi$ in (\ref{N2SYMaction}). In fact
in our calculations
we do not consider diagrams with external matter. Therefore in
(\ref{N2SYMaction}) we keep only the quantum $\Phi$'s and drop
their interactions
with the quantum $V$'s that would never enter in a one-loop diagram.

For the ${\cal
N}=1$ theory we end up with the
sum of three identical ghost loops. Each ghost
contributes to the
one-loop effective action exactly as
a standard background covariant chiral superfield would do, the only
difference being an overall opposite sign because of the statistics.
For the ${\cal N}=2$ theory we obtain contributions from
covariantly chiral loops with a factor
$(-3)$ from the ghosts and a factor
$(+1)$ from the chiral matter.
Thus we explain the general procedure for the computation of a chiral
one-loop contribution. We use a formulation of the background covariant
field method which is particularly simple and it is
valid for real representations of the gauge group as it is the case
of interest to us,
i.e. the adjoint representation. We do not consider matter in complex
representations which are in general affected by anomalies
\cite{anomalies}.

\vspace{0.8cm}

\noindent The rules of background covariant perturbation theory for
commutative supersymmetric Yang-Mills are
clearly explained in ref. \cite{superspace} sect. $6.5$. First we
summarize here the part which is relevant for our specific calculation.
Then we introduce the modifications needed in the noncommutative
case.

In order to obtain the Feynman rules for a general background covariantly
chiral superfield $\Phi$ with action $S_m$
one starts from the generating functional
\beq
Z(J,\bar{J})=\int {\cal D}\Phi~{\cal{D}}\bar{\Phi}~
\exp{\left[S_m~+~(\int d^4x~d^2\theta ~J\Phi~+~{\rm h.c.})\right]}
\label{functint}
\eeq
where both $\Phi$ and $J$ are covariantly chiral $\bar{\nabla}^\ad \Phi=
\bar{\nabla}^\ad J=0$.
The one-loop contribution to the effective action from the $\Phi$
superfield is given by the Gaussian integral over the quantum matter
fields
\beq
\Delta=\int {\cal D}\Phi~{\cal{D}}\bar{\Phi}~
\exp{\left[\int d^4x~d^4\theta ~\bar{\Phi}\Phi\right]}
\label{gaussian}
\eeq
The covariant generalization of the standard method leads to the
introduction of chiral d'Alembertians $\Box_{\pm}$ defined as
\beq
\bar{\nabla}^2 \nabla^2\Phi=\Box_+ \Phi
\qquad\qquad\qquad \nabla^2\bar{\nabla}^2 \bar{\Phi}=\Box_-
\bar{\Phi}
\eeq
with
\bea
&&\Box_+ =\Box-\frac{i}{2}(\nabla^\a {\bf W}_\a)-i{\bf W}^\a\nabla_\a
\nonumber\\
&&\Box_- =\Box-\frac{i}{2}(\bar{\nabla}^\ad \bar{\bf W}_\ad)
-i\bar{\bf W}^\ad\bar{\nabla}_\ad
\label{chiralprop}
\eea
In this way one finds \cite{superspace} that
\beq
\Delta= \exp{\left[\int d^4x~d^2\theta~\frac{1}{2}
\frac{\d}{\d J}[\bar{\nabla}^2\nabla^2-\bar{D}^2D^2]\frac{\d}{\d
J}\right]}~\cdot~\exp{\left[-\int d^4x~d^2\theta~\frac{1}{2}
J\Box_0^{-1}J\right]}
\label{newgaussian}
\eeq
where $\Box_0=1/2 \partial^a \pa_a$ is the free d'Alembertian. A
one-loop graph consists in a string of vertices and propagators, i.e.
a vertex like
$[\bar{\nabla}^2\nabla^2-\bar{D}^2D^2]$, a standard scalar propagator
$(-\Box_0^{-1})$, a vertex again and so on.
The calculation is more conveniently performed using the background
chiral representation,
$\bar{\nabla}_\ad=\bar{D}_\ad$. Indeed this allows to simplify all
the vertices but one, making repeated use of the identity
\beq
[\bar{\nabla}^2\nabla^2-\bar{D}^2D^2]\bar{D}^2=[\Box_+-\Box_0]\bar{D}^2
\label{identity}
\eeq
The net result is that one has to compute one-loop diagrams constructed
with the following propagators and vertices
\bea
&&{\rm propagators:}~~~~~~~~~-\frac{1}{\Box_0}\nonumber\\
&&~~~~~\nonumber\\
&&{\rm all~~~ vertices:}~~~~~\Box_+ -\Box_0 \qquad \qquad\qquad
{\rm but~~~ one:}~~~~~\bar{D}^2(\nabla^2-D^2)
\label{Feynman}
\eea

For a given graph, i.e. for a fixed number of vertices, the completion of
the $D$-algebra is very simple: only one vertex contains $\bar{D}$'s and
the two are both needed in $\d^4(\theta-\theta')\bar{D}^2 D^2
\d^4(\theta-\theta')=\d^4(\theta-\theta')$.
Taking advantage of this covariant technique, the evaluation
of the one-loop contributions from chiral superfields has been reduced to a
rather trivial exercise.

Now we turn to the noncommutative theory and evaluate explicitly the two-
and the three-point function contributions.

\vspace{0.8cm}
\noindent The extension of the above procedure to noncommutative super
Yang-Mills is straightforward \cite{DZ,ASDZ}. One simply has to take
into proper account the $*$-multiplication which
replaces the standard product in the formulas above. This amounts to
the introduction of exponential factors in the vertices, while the
propagators remain unchanged.
In particular at the vertices one has to keep track of the order in
which the quantum fields are Wick contracted, a different order
giving rise to a different exponential factor and ultimately
to a planar or a non planar diagram.
In complete generality a vertex quadratic in the quantum fields can be
written symbolically in momentum space as
\bea
{\cal U}(k_1,k_2,k_3)&=& {\rm Tr}\left( Q(k_1)*{\bf B}(k_2)*Q(k_3)\right)\nonumber\\
&&~~~~\nonumber\\
&=&{\rm Tr}(T^aT^bT^c)~ e^{-\frac{i}{2}(k_1
\times k_2+k_2\times k_3+k_1\times k_3)}~Q_a(k_1){\bf B}_b(k_2)
Q_c(k_3)~~~~~
\label{noncommvert}
\eea
where $Q$ and ${\bf B}$ denote the quantum and the background superfields
respectively. The momenta are flowing into the vertex,
$k_1+k_2+k_3=0$, and we have used the definition $k_i\times k_j\equiv
(k_i)_\m\Theta^{\m\n}(k_j)_\n$ for the $*$-product.
Depending on the order in which the quantum $Q$-fields
are Wick contracted when inserted in the loop, from each vertex
one obtains two types of terms,
an untwisted $P$-term and a twisted $T$-term \cite{DZ,ASDZ}
\bea
&&P\rightarrow {\rm Tr}(T^aT^bT^c)~ e^{-\frac{i}{2}(k_1
\times k_2+k_2\times k_3+k_1\times k_3)}={\rm Tr}(T^aT^bT^c)~
 e^{-\frac{i}{2}k_2\times k_3}\nonumber\\
 &&~~~~\nonumber\\
&&T\rightarrow \pm {\rm Tr}(T^cT^bT^a)~
e^{\frac{i}{2}(k_1
\times k_2+k_2\times k_3+k_1\times k_3)}=\pm {\rm Tr}(T^cT^bT^a)~
e^{\frac{i}{2}k_2\times k_3}
\label{PandT}
\eea
The $+$, $-$ sign in the $T$-term takes into account the bosonic, respectively
fermionic, nature of the background external field.

With all of this in mind, we proceed with the following strategy:
for each diagram the calculation is performed in momentum space, and
first we complete the $D$-algebra in the loop, second we introduce
the exponential factors from the $*$-products at the vertices and
make a distinction between planar and non planar contributions,
finally we evaluate the momentum integral.
In order to accomplish the first step we rewrite the vertices in
(\ref{Feynman}) with flat derivatives and explicit dependence on the
background (the $*$-multiplication between superfields is always
understood)
\bea
\Box_+ -\Box_0& \rightarrow &\Delta_+-i{\bf W}^\a D_\a\nonumber\\
\bar{D}^2(\nabla^2-D^2)&\rightarrow & -i{\bf \Gamma}^\a \bar{D}^2 D_\a
-\frac{1}{2}[i (D^\a {\bf \Gamma}_\a)+{\bf  \G}^\a {\bf  \G}_\a] \bar{D}^2
\label{flatvert}
\eea
where we have defined
\beq
\Delta_+\equiv
-i{\bf  \G}^a\pa_a-\frac{i}{2}(\pa^a{\bf  \G}_a)-\frac{1}{2}{\bf  \G}^a{\bf  \G}_a-\frac{i}{2}(D^\a
{\bf W}_\a)-i{\bf W}^\a {\bf  \G}_\a
\label{D+}
\eeq

Now we go back to (\ref{newgaussian}) and start expanding
the exponentials in order to compute the various contributions to the
one-loop effective action:
\bea
&&{\rm (1):}\qquad\bar{D}^2(\nabla^2-D^2)(-\frac{1}{\Box_0})\nonumber\\
&&~~~\nonumber\\
&&{\rm (2):}\qquad(\Box_+ -\Box_0)(-\frac{1}{\Box_0})
  \bar{D}^2(\nabla^2-D^2)(-\frac{1}{\Box_0})\nonumber\\
&&~~~~~\nonumber\\
&&{\rm (3):}\qquad(\Box_+ -\Box_0)(-\frac{1}{\Box_0})(\Box_+
  -\Box_0)(-\frac{1}{\Box_0})
  \bar{D}^2(\nabla^2-D^2)(-\frac{1}{\Box_0})\nonumber\\
&&~~~~~\nonumber\\
&&{\rm (4):}\qquad(\Box_+ -\Box_0)(-\frac{1}{\Box_0})
(\Box_+ -\Box_0)(-\frac{1}{\Box_0})(\Box_+
-\Box_0)(-\frac{1}{\Box_0})
  \bar{D}^2(\nabla^2-D^2)(-\frac{1}{\Box_0})\nonumber\\
&&~~~~~\nonumber\\
&&\qquad\qquad\dots\dots\dots
\label{expansion}
\eea
Since at least two $D$'s and two $\bar{D}$'s are needed for a nonzero
result, from (\ref{flatvert}) we immediately obtain that the first
order expansion does not contribute. Moreover since the vertex
$(\Box_+ -\Box_0$) contains at most one $D$ that can be used for the
$D$-algebra, the second order expansion only contributes to the
two-point function
\bea
(\Box_+ -\Box_0)(-\frac{1}{\Box_0})
  \bar{D}^2(\nabla^2-D^2)(-\frac{1}{\Box_0})&\rightarrow&
  -i{\bf W}^\a D_\a(-\frac{1}{\Box_0})(-i{\bf \Gamma}^\b \bar{D}^2 D_\b)
  (-\frac{1}{\Box_0})\nonumber\\
  &&~~~~\nonumber\\
  &\rightarrow & {\bf W}^\a (-\frac{1}{\Box_0}){\bf  \G}_\a (-\frac{1}{\Box_0})
\label{2point}
\eea
According to the general strategy, having completed the $D$-algebra
we have to take into account the nature of the quantum background
vertices we started with. Going back to (\ref{PandT}), we learn that
the combination $PP+TT$ corresponds to the planar graphs, while
$PT+TP$ gives rise to the non planar ones.
Thus we obtain the following contributions to the one-loop two-point
function:
\bea
&&  \G^{(2)}_{\rm planar}=~N~\int d^2\theta~\frac{d^4p}{(2\p)^4}~
{\rm Tr}\left({\bf W}^\a(p){\bf W}_\a(-p)\right)~~~~~~~~~~~~~~~~~~\nonumber\\
&&~~~~~~~~~~~~~~~~~~~~~~~~~~~~~~~~~~~\int \frac{d^4k}{(2\p)^4}~
\frac{1}{(k^2+m^2)[(p+k)^2+m^2]}
\label{2pointplanar}
\eea
and
\bea
&&  \G^{(2)}_{\rm nonplanar}=~-~\int d^2\theta~\frac{d^4p}{(2\p)^4}~
{\rm Tr}\left({\bf W}^\a(p)\right){\rm Tr}\left({\bf
W}_\a(-p)\right)~~~~~~~~~~~~~~~~~~~\nonumber\\
&&~~~~~~~~~~~~~~~~~~~~~~~~~~~~~~~~~~~~~~~~~\int \frac{d^4k}{(2\p)^4}~
\frac{e^{ip\times k}}{(k^2+m^2)[(p+k)^2+m^2]}
\label{2pointnonplanar}
\eea
where $m$ is an IR mass regulator.
The result in (\ref{2pointplanar}) reproduces the
logarithmic UV divergence that leads to a nonvanishing one-loop
$\b$-function  both for the ${\cal N}=1$ and the ${\cal N}=2$ theories,
with the counting that we already mentioned, i.e. $(-3)$ times the
contribution in (\ref{2pointplanar}) from the three ghosts in the
${\cal N}=1$ case, $(-3+1)$ from the three ghosts and one chiral
matter for the ${\cal N}=2$ case.

The corresponding nonplanar result in (\ref{2pointnonplanar}) is
finite in the UV thanks to the presence of the non local scale
$\D x^\m=\Theta^{\m\n}p_\n$ that acts as a short distance cutoff. In
the low-energy approximation, $p_i\cdot p_j$ small,
with $p_i\times p_j$ finite it is easy to perform the $k$-integration.
With the definition
$p\circ p=p_\m\Theta^{\m\n}\Theta_{\rho\n}p^{\rho}$, one obtains
\beq
  \G^{(2)l.e.}_{\rm nonplanar}
=~-\frac{1}{8\p^2}~\int d^2\theta~\frac{d^4p}{(2\p)^4}~
{\rm Tr}\left({\bf W}^\a(p)\right){\rm Tr}\left({\bf
W}_\a(-p)\right)~K_0(m\sqrt{p\circ p})
\label{2pointnonplanarle}
\eeq
Since
\beq
\lim_{z\rightarrow 0} ~K_0(z)= -\ln{\frac{z}{2}}
\eeq
the expression in (\ref{2pointnonplanarle}) is divergent in the IR
\cite{perturb}. In a sense the real novelty of noncommutative
theories is given by the interplay between the UV and the IR
behaviour. The lack of conformal invariance of the quantum commutative
theory
($\b \neq 0$) leads to a nonperturbative, singular behaviour in the non
local scale ($\Theta \neq 0$).
A complete understanding of the role played by this effect
is likely to be found from string dynamics \cite{IR}.

Now we continue in our task and turn to the computation of the
vector three-point function due to a chiral matter loop.

\vspace{0.8cm}

\noindent As observed before,
the second order expansion in (\ref{expansion})
contributes only to the two-point function. Thus
the three-point function is obtained from the third order terms
\beq
(\Box_+ -\Box_0)(-\frac{1}{\Box_0})(\Box_+
  -\Box_0)(-\frac{1}{\Box_0})
  \bar{D}^2(\nabla^2-D^2)(-\frac{1}{\Box_0})
\label{third}
\eeq
where at every vertex we only keep terms which are linear in the
background fields.
Inserting (\ref{flatvert}) and (\ref{D+}) in (\ref{third}) we obtain
several contributions, all with trivial $D$-algebra, that we list:
\bea
{\rm (a):}&&(-i{\bf W}^\g D_\g)~(-\frac{1}{\Box_0})~
(-i{\bf W}^\b D_\b)~(-\frac{1}{\Box_0})~\bar{D}^2 (-\frac{i}{2}
D^\a {\bf  \G}_\a)~(-\frac{1}{\Box_0})\nonumber\\
&&~~~~\longrightarrow~~~{\bf W}^\b~(-\frac{1}{\Box_0})~
{\bf W}_\b~(-\frac{1}{\Box_0})~
(-\frac{i}{2}D^\a {\bf  \G}_\a)~(-\frac{1}{\Box_0})\nonumber\\
&&~~~~~\nonumber\\
{\rm (b):}&&(-i{\bf W}^\g D_\g)~(-\frac{1}{\Box_0})~
(-i{\bf W}^\b D_\b)~(-\frac{1}{\Box_0})~ (-i{\bf  \G}^\a
\bar{D}^2 D_\a )~(-\frac{1}{\Box_0})\nonumber\\
&&~~~~\longrightarrow~~~(-{\bf W}^\g)~(-\frac{1}{\Box_0})~
(D_\g{\bf W}^\b)~(-\frac{1}{\Box_0})~
(i{\bf  \G}_\b)~(-\frac{1}{\Box_0})\nonumber\\
&&~~~~~~~~~~~~+{\bf W}^\b~(-\frac{1}{\Box_0})~
{\bf W}_\b~(-\frac{1}{\Box_0})~
(iD^\a {\bf  \G}_\a)~(-\frac{1}{\Box_0})\nonumber\\
&&~~~~~\nonumber\\
{\rm (c):}&&[-i{\bf  \G}^a\pa_a-\frac{i}{2}(\pa^a{\bf  \G}_a)
-\frac{i}{2}D^\a{\bf
W_\a}]~(-\frac{1}{\Box_0})~
(-i{\bf W}^\b D_\b)~(-\frac{1}{\Box_0})~ (-i{\bf  \G}^\g
\bar{D}^2 D_\g )~(-\frac{1}{\Box_0})\nonumber\\
&&~~~~\longrightarrow~~~[-i{\bf  \G}^a\pa_a-\frac{i}{2}(\pa^a{\bf  \G}_a)
-\frac{i}{2}D^\a{\bf W_\a}]~(-\frac{1}{\Box_0})~
{\bf W}^\b~(-\frac{1}{\Box_0})~
{\bf  \G}_\b~(-\frac{1}{\Box_0})\nonumber\\
&&~~~~~\nonumber\\
{\rm (d):}&&(-i{\bf W}^\b D_\b)~(-\frac{1}{\Box_0})~
[-i{\bf  \G}^a\pa_a-\frac{i}{2}(\pa^a{\bf  \G}_a)-\frac{i}{2}D^\a{\bf
W_\a}]~(-\frac{1}{\Box_0})~(-i{\bf  \G}^\g
\bar{D}^2 D_\g )~(-\frac{1}{\Box_0})\nonumber\\
&&~~~~\longrightarrow~~~{\bf W}^\b~(-\frac{1}{\Box_0})~
[-i{\bf  \G}^a\pa_a-\frac{i}{2}(\pa^a{\bf  \G}_a)
-\frac{i}{2}D^\a{\bf W_\a}]~(-\frac{1}{\Box_0})~{\bf  \G}_\b
~(-\frac{1}{\Box_0})\nonumber\\
&&~~~~~
\label{3point}
\eea
Having completed the $D$-algebra now we have to take into account the
$*$-products and the orderings from Wick contractions at the vertices.
We have external momenta $p_1$, $p_2$, $p_3$ with $p_1+p_2+p_3=0$ and
a momentum $k$ in the loop. Planar diagrams correspond to
arrangements of the vertices like $P_1P_2P_3$ and $T_1T_2T_3$, while
non planar diagrams correspond to the six terms
$P_1P_2T_3$, $P_1T_2P_3$,
$T_1P_2P_3$, $T_1T_2P_3$, $T_1P_2T_3$, $P_1T_2T_3$. In order
to write the final result in a simple form it is convenient to
arrange the external background fields in the most symmetric way,
i.e. for every term we write
\beq
{\bf A}_a(p_1)~{\bf B}_b(p_2)~{\bf C}_c(p_3)~+~ {\rm all~ permutations ~~
(p_1a)\leftrightarrow
(p_2b)\leftrightarrow (p_3c)}
\label{extback}
\eeq
Now we are ready to analyze the contributions in (\ref{3point}).
We can use the symmetry property in (\ref{extback}) and exchange
the external fields (keeping track of the correct sign when
fermionic fields are involved). Also we can integrate by parts
derivatives in order to reduce the various terms to similar
structures. In this way one finds immediately
that the contributions containing the
connection ${\bf \G}_a$ cancel. In fact making use of the relation
\beq
D_\a {\bf W}_\b= \frac{1}{2} D_{(\a}{\bf W}_{\b)}+\frac{1}{2}C_{\b\a}
D^\g {\bf W}_\g
\eeq
it is easy
to show that also the remaining terms add up to zero.
Thus the chiral matter contribution to the vector three-point
function is zero. This result, combined with the fact that vectors
circulating in the loop do not contribute (not enough $D$'s in the loop),
 leads to the
conclusion that the one-loop three-point function with vector external lines
is zero for ${\cal N}=1$, ${\cal N}=2$ and ${\cal N}=4$
supersymmetric Yang-Mills. The simplicity of the above calculation
is a good example of the power of the background covariant superfield approach
as compared to standard superfield techniques (cf. \cite{CZ}) or
to component field methods.

\vspace{0.8cm}

\noindent Finally we consider the chiral loop corrections to the
effective action with four vector external fields.
Here we explain the general procedure while the details and the
presentation of the complete answer will be given elsewhere
\cite{dan}.

One obtains contributions from the third order expansion in (\ref{expansion})
with one
of the vertices quadratic in the background, and contributions from
the fourth order expansion with all the vertices linear in the
background. It is straightforward to list the various terms and
perform the $D$-algebra, exactly as we have done in the previous two
examples. Once again one can exploit the symmetries of the four-point
function, i.e. for every term we
write
\bea
&&{\bf A}_a(p_1)~{\bf B}_b(p_2)~{\bf C}_c(p_3)~{\bf D}_d(p_4)~+~
{\rm all~ permutations ~~
(p_1a)\leftrightarrow
(p_2b)\leftrightarrow (p_3c)\leftrightarrow (p_4d)}~\nonumber\\
&&~~~~
\label{4extback}
\eea
In this way the result is drastically simplified and
one can isolate the terms which, at the component level, give rise to
structures with four external $F_{\m\n}$ field strengths. To this end
one has to remember that the bosonic field strength $F_{\m\n}$ is contained
in the superfield \cite{superspace}
\beq
D_{(\a}{\bf W}_{\b)}\rightarrow f_{\a\b}=\s^{\m\n}_{\a\b}F_{\m\n}
\eeq
where $(\s_{\m\n})_\a^{~\b}\equiv-1/4(\s_\m \tilde{\s}_\n-\s_\n
 \tilde{\s}_\m)_\a^{~\b}$
and $(\s_\m)_{\a\ad}\equiv(1,\vec{\s})$,
$(\tilde{\s}_\n)^{\ad\a}\equiv(1,-\vec{\s})$.
Back to the four-point function, one finds  relevant contributions
in the form
\beq
{\bf W}_\g~(-\frac{1}{\Box_0})~(D^\b {\bf W}^\g)~(-\frac{1}{\Box_0})~
(D_\b {\bf W}^\a)~
(-\frac{1}{\Box_0})~{\bf\G}_\a (-\frac{1}{\Box_0})
\label{4point}
\eeq
The external fields, completely symmetrized, are arranged as follows
\bea
&&{\cal R}(1a,2b,3c,4d)\equiv{\bf W}_{a \g}(p_1)(D^\b {\bf W}_b^\g)(p_2)
(D_\b {\bf W}_c^\a)(p_3){\bf\G}_{d\a}(p_4)+~~~~~~~~~~\nonumber\\
&&~~~~\nonumber\\
&&~~~~~~~~~~~~~~~~~~~~~+~{\rm all~ permutations ~~
(p_1a)\leftrightarrow
(p_2b)\leftrightarrow (p_3c)\leftrightarrow (p_4d)}
\label{4pointback}
\eea
so that the above expression can be compared directly with the bosonic
component amplitude since
\beq
\int d^4\theta ~{\cal R}(1a,2b,3c,4d)~\rightarrow~
t^{\m_1\n_1\m_2\n_2\m_3\n_3\m_4\n_4} F_{\m_1\n_1}^a(p_1)F_{\m_2\n_2}^b(p_2)
F_{\m_3\n_3}^c(p_3)F_{\m_4\n_4}^d(p_4)
\label{boschiral}
\eeq
where $t^{\m_1\n_1\m_2\n_2\m_3\n_3\m_4\n_4}$ is the symmetric tensor
given e.g. in (9.A.18) of \cite{GSW}.

In addition to terms like the ones in (\ref{4point}), one obtains
also terms in the form
\beq
{\bf W}^\a~(-\frac{1}{\Box_0})~{\bf W}_\a~(-\frac{1}{\Box_0})~
\bar{{\bf W}}^\ad~
(-\frac{1}{\Box_0})~\bar{{\bf W}}_\ad (-\frac{1}{\Box_0})
\label{4point2}
\eeq
which are generated by the appearance of structures like
$i\pa_a {\bf \G}^\a-iD^\a{\bf \G}_a= 2\bar{\bf W}_\ad$.
 Using the relation
\bea
&&\int d^4\theta ~~\left[ {\bf W}^{\a a}(p_1){\bf W}_\a^b(p_2)
\bar{{\bf W}}^{\ad c}(p_3)
\bar{{\bf W}}_\ad^d(p_4)\right.~~~~~~~~~~~~~~~~~~~~~~~~\nonumber\\
&&~~~~\nonumber\\
&&~~~~~~~~~~~~~~~~~~~~~\left.+~{\rm all~ permutations ~~
(p_1a)\leftrightarrow
(p_2b)\leftrightarrow (p_3c)\leftrightarrow (p_4d)}\right]\nonumber\\
&&~~~~~~\nonumber\\
&&\rightarrow
t^{\m_1\n_1\m_2\n_2\m_3\n_3\m_4\n_4} F_{\m_1\n_1}^a(p_1)F_{\m_2\n_2}^b(p_2)
F_{\m_3\n_3}^c(p_3)F_{\m_4\n_4}^d(p_4)
\label{bosvector}
\eea
 from the
superfield result in (\ref{4point2}) once again one can extract the bosonic
contributions.

Then in all the various terms
one has to  insert the appropriate $*$-products that give rise
to planar and non planar diagrams, exactly as it has been done in
\cite{DZ,ASDZ}.
At that stage one evaluates the loop
momentum integrals with four propagators
\beq
I_0(k;p_1,\dots,p_4)=\frac{1}{[(k+p_1)^2+m^2][ k^2+m^2][ (k-p_4)^2+m^2]
[ (k+p_1+p_2)^2+m^2]}
\label{box}
\eeq
for each diagram.

Once all the terms have been collected,
again as we have argued for the two-point function, we have to multiply
the final answer by $(-3)$ from the three
chiral ghosts in the case of the ${\cal N}=1$
theory , while for
${\cal N}=2$ we have the same contribution times the factor
$(-3+1)$ which counts the three ghosts and one chiral
superfield. The complete four-point function with vector external
fields \cite{dan} is then obtained by adding to the one-loop matter corrections
the corresponding one-loop vector corrections computed in \cite{ASDZ}.
It would be interesting to evaluate the corresponding amplitudes
in the appropriate $D3$-brane configurations and compare the field
theory limit from the string approach to the one-loop field theory effective
action calculations.

A neat result that we have obtained is the vanishing of the
vector three-point function in all ${\cal N}=1,2,4$ supersymmetric
Yang-Mills theories. Since we have used the background field quantization
this result is a gauge invariant statement. It is in correspondence
with the fact that
in the supersymmetric Born-Infeld action terms cubic in the
$F_{\m\n}$ field strength are absent.

The superfield background covariant
method provides a very efficient mean of computing in perturbation
theory. One could take advantage of this
and study renormalization properties
of correlators of gauge invariant operators in noncommutative
field theories \cite{GHI,IKK}.

\vspace{1.5cm}

\noindent
{\bf Acknowledgements}

\noindent
I thank Marc Grisaru and Carlos Nunez for interesting conversations and
the Department of Harvard University for warm hospitality while
this work has been performed.
This work has been partially supported by INFN, MURST, and the
European Commission RTN program HPRN-CT-2000-00113 in which I
am associated to the University of Torino.


\newpage
\begin{thebibliography}{99}
\bibitem{all} A. Connes, M.R. Douglas, A. Schwarz, JHEP {\bf 02 }
(1998) 003, hep-th/9711162\\
M.R. Douglas, C. Hull, JHEP {\bf 02} (1998) 008, hep-th/9711165
\bibitem{SW} N. Seiberg, E. Witten, JHEP {\bf 09} (1999) 032, hep-th/9908142
\bibitem{others} N. Seiberg, L. Susskind, N. Toumbas, JHEP {\bf 06} (2000) 021,
hep-th/000540\\
R. Gopakumar, J. Maldacena, S. Minwalla, A. Strominger, JHEP {\bf 06}
(2000) 036, hep-th/0005048\\
R. Gopakumar,S. Minwalla, N. Seiberg, A. Strominger, JHEP {\bf 08}
(2000) 008, hep-th/0006062\\
E. Bergshoeff, D.S. Berman, J.P. van der Schaar, P. Sundell,
Phys. Lett. {\bf B492} (2000) 193, hep-th/0006112
\bibitem{string} O. Andreev, H. Dorn, Nucl. Phys. {\bf B583}
(2000) 145, hep-th/0003113\\
A. Bilal, C.-S. Chu, R. Russo, Nucl. Phys. {\bf B582} (2000) 65,
hep-th/0003180\\
C.-S. Chu, R. Russo, S. Sciuto, Nucl.Phys. {\bf B585} (2000) 193,
hep-th/0004183
\bibitem{perturb} T. Filk, Phys. Lett. {\bf B376}(1996) 53\\
 S. Minwalla, M. Van Raamsdonk, N. Seiberg, hep-th/9912072\\
A. Matusis, L. Susskind, N. Toumbas, hep-th/0002075\\
M. Van Raamsdonk, N. Seiberg, JHEP {\bf 03} (2000) 035, hep-th/0002186\\
M.M. Sheikh-Jabbari, JHEP {\bf 9906} (1999) 015, hep-th/9903107\\
C.P. Martin, D. Sanchez-Ruiz, Phys. Rev. Lett. {\bf 83} (1999) 476,
hep-th/9903077\\
T. Krajewski, R. Wulkenhaar, J. Mod. Phys. {\bf A15} (2000) 1011,
hep-th/9903187\\
I.Y. Arefeva, D.M. Belov, A.S. Koshelev, O.A. Rychkov, Phys. Lett.
{\bf B487} (2000) 357, hep-th/0003176\\
C.P. Martin, F. Ruiz Ruiz, hep-th/0007131\\
I. Chepelev, R. Roiban, JHEP {\bf 05} (2000) 037, hep-th/9911098;
hep-th/0008090\\
M. Haykawa, Phys. Lett. {\bf B478} (2000) 394, hep-th/9912094;
hep-th/9912167\\
A. Armoni, hep-th/0005208\\
V.V. Khoze, G. Travaglini, hep-th/0011218
\bibitem{DZ} D. Zanon, hep-th/0009196
\bibitem{ASDZ} A. Santambrogio, D. Zanon, hep-th/0010275\\
M. Pernici, A. Santambrogio, D. Zanon, hep-th/0011140
\bibitem{FL}  S. Ferrara, M.A. Lledo', JHEP {\bf 05} (2000) 008, hep-th/0002084
\bibitem{GS} M.T. Grisaru, W. Siegel, Nucl. Phys. {\bf B201} (1982)
292
\bibitem{GS2} M.T. Grisaru, W. Siegel, Phys. Lett. {\bf 110B} (1982)
49
\bibitem{superspace} S.J. Gates, M.T. Grisaru, M. Rocek, W. Siegel,
``Superspace" (Benjamin-Cummings, Reading, MA, 1983)
\bibitem{*trek} H. Liu, J. Michelson, hep-th/0008205
\bibitem{GZ} M.T. Grisaru, D. Zanon, Nucl. Phys {\bf B252} (1985) 578
\bibitem{anomalies} L. Bonora, M. Schnabl, M.M. Sheikh-Jabbari, A. Tomasiello,
Phys. Lett. {\bf B485} (2000) 311, hep-th/0002210; Nucl. Phys. {\bf B589}
(2000) 461, hep-th/0006091\\
C.P. Martin, hep-th/0008126\\
F. Ardalan, N. Sadooghi, hep-th/0009233\\
M.T. Grisaru, S. Penati, hep-th/0010177
\bibitem{IR} H. Liu, J. Michelson, Phys. Rev. {\bf D62} (2000) 066003,
hep-th/0004013
\bibitem{CZ} W.E. Caswell, D. Zanon, Nucl. Phys. {\bf B182} (1981) 125
\bibitem{dan} D. Zanon, in preparation
\bibitem{GSW} M.B. Green, J.H. Schwarz, E. Witten, ``Superstring Theory"
(Cambridge University Press, 1987)
\bibitem{GHI} S. Das, S.-J. Rey, hep-th/0008042\\
D.J. Gross, A. Hashimoto, N. Itzhaki, hep-th/0008075\\
T. Mehen, M.B. Wise, hep-th/0010204\\
H. Liu, hep-th/0011125
\bibitem{IKK} N. Ishibashi, S. Iso, H. Kawai, Y. Kitazwa,
Nucl. Phys. {\bf B573} (2000) 573, hep-th/9910004


\end{thebibliography}
\end{document}